\documentclass[pdflatex,sn-nature,oneside]{sn-jnl}

\usepackage{graphicx}%
\usepackage{multirow}%
\usepackage{amsmath,amssymb,amsfonts}%
\usepackage{amsthm}%
\usepackage{mathrsfs}%
\usepackage[title]{appendix}%
\usepackage{xcolor}%
\usepackage{textcomp}%
\usepackage{manyfoot}%
\usepackage{booktabs}%
\usepackage{algorithm}%
\usepackage{algorithmicx}%
\usepackage{algpseudocode}%
\usepackage{listings}%
\usepackage{float}
\usepackage{hyperref}
\usepackage{caption}

\theoremstyle{thmstyleone}%
\theoremstyle{thmstyletwo}%

\theoremstyle{thmstylethree}%

\begin{document}

\title[Article Title]{scpFormer: A Foundation Model for Unified Representation and Integration of the Single-Cell Proteomics}


\author[1]{\fnm{Qifeng} \sur{Zhou}}\email{qxz8706@mavs.uta.edu}
\equalcont{These authors contributed equally to this work.}

\author[2]{\fnm{Lei} \sur{Yu}}\email{Lei.Yu@UTSouthwestern.edu}
\equalcont{These authors contributed equally to this work.}

\author[1]{\fnm{Yuzhi} \sur{Guo}}\email{yuzhi.guo@mavs.uta.edu}

\author[1]{\fnm{Yuwei} \sur{Miao}}\email{yxm9326@mavs.uta.edu}

\author[1]{\fnm{Hehuan} \sur{Ma}}\email{hehuan.ma@mavs.uta.edu}

\author[1]{\fnm{Wenliang} \sur{Zhong}}\email{wxz9204@mavs.uta.edu}

\author*[2]{\fnm{Lin} \sur{Xu}}\email{Lin.Xu@UTSouthwestern.edu}

\author*[1]{\fnm{Junzhou} \sur{Huang}}\email{jzhuang@uta.edu}

\affil[1]{\orgdiv{Department of Computer Science and Engineering}, \orgname{The University of Texas at Arlington}, \orgaddress{\city{Arlington}, \postcode{76010}, \state{TX}, \country{USA}}}

\affil[2]{\orgdiv{Quantitative Biomedical Research Center, Department of Health Data Science and Biostatistics, Peter O’Donnell Jr. School of Public Health}, \orgname{University of Texas Southwestern Medical Center}, \orgaddress{\city{Dallas}, \postcode{75390}, \state{TX}, \country{USA}}}

\abstract{The integration of single-cell proteomic data is often hindered by the fragmented nature of targeted antibody panels. To address this limitation, we introduce scpFormer, a transformer-based foundation model designed for single-cell proteomics. Pre-trained on over 390 million cells, scpFormer replaces standard index-based tokenization with a continuous, sequence-anchored approach. By combining Evolutionary Scale Modeling (ESM) with value-aware expression embeddings, it dynamically maps variable panels into a shared semantic space without artificial discretization. We demonstrate that scpFormer generates global cell representations that perform competitively in large-scale batch integration and unsupervised clustering. Moreover, its open-vocabulary architecture facilitates in silico panel expansion, assisting in the reconstruction of biological manifolds in sparse clinical datasets. Finally, this learned protein co-expression logic is transferable to bulk-omics tasks, supporting applications like cancer drug response prediction. scpFormer provides a versatile, panel-agnostic framework to facilitate scalable biomarker discovery and precision oncology.}

\maketitle
\section{Introduction}
Single-cell proteomics (SCP) sequencing technologies have been transforming our understanding of tissue organization, developmental trajectories, and disease mechanisms~\cite{perkel2021single, kelly2020single, bennett2023single, mund2022deep, guldberg2023computational}. Technological innovations over the past decade have enabled protein measurements at single-cell resolution through complementary platforms~\cite{perkel2021single, bennett2023single, slavov2022counting, mund2022deep, petelski2021multiplexed, truong2024s, ye2025enhanced, gatto2023initial}, including CITE-seq, CyTOF, and others. These technologies have revealed functional heterogeneity invisible to transcriptomics alone, particularly in immune and cancer systems~\cite{mali2023single, fernandez2019single, li2024proteomics, rosenberger2023spatial}. Although single-cell RNA sequencing (scRNA-seq) remains one of most popular methods for dissecting cellular heterogeneity, transcriptomic measurements primarily capture transcriptional potential rather than the executed functional state of cells~\cite{furtwangler2025mapping}. Proteins, through their abundance, localization, post-translational modifications, and interaction dynamics, constitute the direct effectors of biological function, governing signaling cascades, enzymatic activity, and structural organization. Extensive evidence demonstrates that mRNA abundance correlates at best modestly with protein levels~\cite{furtwangler2025mapping}. Consequently, key functional determinants, such as surface marker abundance, phosphorylation states, and cytokine secretion, are frequently decoupled from their transcriptomic proxies. This disconnect is especially consequential in clinical and translational studies, such as immunotherapy response, tumor immune evasion, and drug resistance, where immediate protein-level events at single-cell resolution closely determine cell fate, while reliance on RNA proxies may obscure these studies. Consequently, single-cell proteomics has emerged as an indispensable modality for resolving biological mechanisms that remain opaque to RNA-centric analyses~\cite{perkel2021single, guldberg2023computational, wu2026single, gatto2023initial}.

In parallel, technological innovations in generative artificial intelligence, most notably Transformer-based foundation models, have revolutionized the analysis of high-dimensional biological data. The Transformer architecture has emerged as a dominant framework for modeling complex biological systems, utilizing self-attention mechanisms to capture long-range dependencies within cellular networks~\cite{vaswani2017attention, wang2023scientific, moor2023foundation}. In single-cell transcriptomics, foundation models such as scGPT and Geneformer have demonstrated strong performance in cell state inference, perturbation modeling, and regulatory prediction.~\cite{cui2024scgpt, theodoris2023transfer}. The development of DNA language models~\cite{ji2021dnabert, dalla2025nucleotide, nguyen2023hyenadna, sanabria2024dna} based on utilizing DNA sequence information also brings a series of technological innovations to the genomics and epigenomics fields. Similarly, protein language models trained on amino-acid sequences (e.g., ESM~\cite{lin2023evolutionary, hayes2025simulating} and AlphaFold~\cite{jumper2021highly, abramson2024accurate}) have shown that evolutionary information embedded in sequences can be leveraged to predict structure and function with notable accuracy. 

Despite these advances, a foundation model for single-cell proteomics data has not yet been established. Current single-cell proteomics analyses predominantly rely on traditional statistical workflows or ad hoc adaptations of bulk proteomics pipelines~\cite{zhao2025deciphering, truong2024s}, which treat proteins as independent features and fail to capture higher-order co-expression and functional dependencies at the cellular level. To date, no foundation models have been developed to learn a unified representation of the single-cell proteome across technologies, tissues, and experimental designs.

The absence of a foundation model for single-cell proteomics is not incidental but reflects a set of modality-specific challenges that distinguish a single-cell proteomics expression model from DNA and protein sequence-based language models~\cite{jumper2021highly, lin2023evolutionary} and single-cell transcriptomics expression foundation models~\cite{cui2024scgpt, hao2024large}. First, unlike DNA and RNA, which can be enzymatically amplified by polymerase chain reaction (PCR) in next-generation sequencing (NGS) workflows, proteins cannot be sequence-specifically amplified, requiring direct detection of native molecules and imposing fundamental physical limits on sensitivity that lead to systematic, non-random missingness in single-cell proteomics. As a result, protein measurements exhibit structured patterns of missing values that reflect underlying molecular abundance and chemistry, rather than random noise, posing significant challenges for large language models that typically assume dense or randomly corrupted inputs. Second, SCP data are characterized by extreme sparsity and noise, arising from stochastic ion sampling in mass spectrometry or non-specific binding and background in antibody-based assays, which undermines assumptions underlying standard imputation and denoising strategies. Third and most critically, single-cell proteomics suffers from feature discontinuity and fragmentation. Unlike scRNA-seq, which benefits from a universal and stable characterization of ~approximately 20,000 protein-coding genes, SCP datasets are inherently heterogeneous with respect to the number of proteins measured across studies: many experiments rely on bespoke antibody panels with minimal overlap, while mass spectrometry-based experiments exhibit variable protein coverage due to stochastic sampling. This fragmentation exposes a fundamental limitation of standard Transformer architectures: their reliance on a fixed vocabulary. In conventional implementations, each protein is treated as a discrete token with a static, learnable index from a closed dictionary established during pre-training. Although this approach suffices for modalities with a consistent reference, it is fundamentally ill-suited for the open-ended nature of single-cell proteomics. Because fixed-vocabulary models cannot process proteins outside their pre-defined training set, researchers are compelled to discard unique, non-overlapping markers. This closed-set restriction creates severe data silos, preventing the integration of disparate datasets and limiting the capacity of traditional architectures to capture the heterogeneous landscape of the single-cell proteome.

To address these fundamental limitations, we present scpFormer, a transformer-based generative foundation model designed specifically to decode the single-cell proteome. Based on the biophysical realities of single-cell proteomics, we replace rigid, dataset-bound feature indexing with a continuous, panel-agnostic tokenization strategy, coupled with a pre-training framework engineered explicitly for the continuous physiological gradients of protein expression. This framework empowers scpFormer with two distinct operational capacities: robust representation extraction and context-aware generation. Through comprehensive benchmarking, we demonstrate that the global cell representations derived from scpFormer perform competitively in unsupervised clustering and large-scale batch integration, addressing complex technical confounders while preserving biological boundaries. Furthermore, we highlight its generative capabilities by facilitating in silico protein imputation and assisting in the reconstruction of biological manifolds in sparse clinical cohorts. Finally, demonstrating its translational utility, we show that scpFormer’s single-cell derived proteomic embeddings can generalize to bulk-omics tasks, supporting the prediction of cancer drug responses. Collectively, this work establishes the first foundation-model framework for single-cell proteomics, providing a scalable and panel-agnostic computational substrate for future biomarker discovery and precision oncology.

\section{Results}
\subsection{Overview of scpFormer}

\begin{figure}[htbp]
  \centering
  \includegraphics[width=\textwidth]{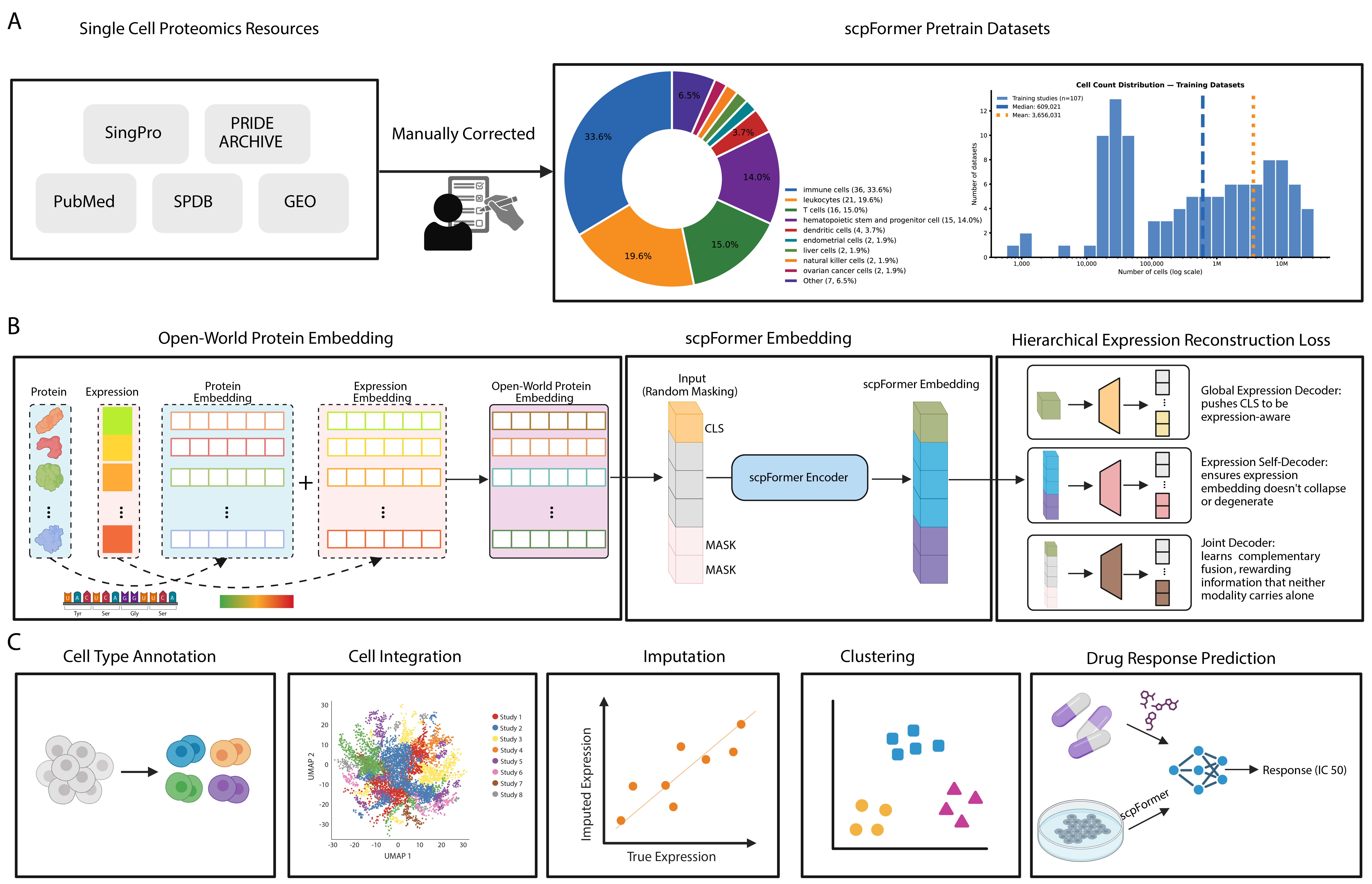} 
  \caption{\textbf{Overview of the scpFormer framework for single-cell proteomics.} \textbf{A,} Large-scale data curation and pre-training corpus construction. Single-cell proteomics datasets were aggregated from diverse public repositories (SingPro, PRIDE, PubMed, SPDB, and GEO) and subjected to rigorous manual correction. The resulting large-scale pre-training corpus encompasses highly heterogeneous cellular phenotypes, as detailed by the cell-type composition (pie chart) and the extensive cell count distribution (histogram). \textbf{B,} Model architecture and pre-training strategy. scpFormer utilizes an Open-World Protein Embedding strategy (Identity-Abundance Decoupled Tokenization, IADT) to process single-cell inputs. Continuous expression values and corresponding continuous protein identity embeddings (derived from amino acid sequences) are mapped into a shared dense space and fused. Following dynamic random masking, the sequence, prepended with a learnable [CLS] token, is processed by the scpFormer encoder. The network is jointly optimized via a Hierarchical Expression Reconstruction framework, which includes an Expression Self-Decoder, a Global Expression Decoder, and a Joint Decoder. \textbf{C,} Downstream analytical tasks. The pre-trained universal cellular representation serves as a robust information bottleneck, seamlessly adapting to diverse downstream applications including cell type annotation, multi-batch cell integration, missing value imputation, clustering, and multi-modal drug response prediction.}
  \label{fig:fig_overview}
\end{figure}

We present scpFormer, a transformer-based generative foundation model for decoding single-cell proteomics. Trained on over 390 million cells across highly heterogeneous experimental platforms, including mass cytometry, CITE-seq, and multiplexed imaging, scpFormer encompasses more than 530 proteins within a 21-million-parameter architecture (Fig. \ref{fig:fig_overview}A). As the first foundation model purpose-built for single-cell proteomics, scpFormer introduces two key innovations tailored to the unique challenges of single cell proteomic data.

The first innovation addresses feature fragmentation across proteomics inputs. Unlike transcriptomics, where gene vocabularies are largely standardized, proteomic studies routinely employ disjoint, highly targeted antibody panels with no universal reference space for cross-experiments. scpFormer overcomes this by replacing discrete token indices with a continuous semantic tokenization strategy: leveraging a pre-trained Evolutionary Scale Modeling (ESM) encoder~\cite{hayes2025simulating}, each measured protein is projected into a shared structural and functional manifold derived from its canonical amino acid sequence. This decouples protein identity from arbitrary dataset-specific indices, enabling scpFormer to process variable-length, panel-agnostic inputs and generalize zero-shot to proteins absent from pre-training. Complementing this, scpFormer adopts a continuous value-aware embedding strategy. Rather than discretizing expression values which compresses dynamic range and erases subtle phenotypic gradients raw, non-zero expression magnitudes are explicitly projected into a high-dimensional space and dynamically fused with their ESM-derived semantic identities. This dual-modality embedding, termed Open-World Protein Embedding, enables the self-attention mechanism to jointly encode both the identity of each protein and its expression level (Fig. \ref{fig:fig_overview}B).

The second innovation captures the complexity of protein regulatory networks through a hierarchical pre-training framework. As illustrated in (Fig. \ref{fig:fig_overview}B), scpFormer is pre-trained via a unified, single-iteration multi-objective framework tailored to single-cell proteomic data. Rather than relying on a simple reconstruction task, we optimize three synergistic generative decoders. First, the Expression Self-Decoder captures local, micro-level protein-protein interactions. It imputes missing target proteins based exclusively on the observed co-expression network of visible proteins, ensuring the baseline expression embeddings do not collapse or degenerate. Second, the Global Expression Decoder shifts the focus to the macro-level cellular state. It explicitly pushes the global classification token ([CLS]) to be fully "expression-aware," compressing the entire cellular phenotype into a robust, unified anchor that can independently predict missing protein abundances. Finally, the Joint Decoder bridges the micro and macro scales. By fusing the mature global cellular state with individual protein identities, it learns complementary regulatory information that neither local context nor global state carries alone. By optimizing these three decoders concurrently, scpFormer is forced to learn deeply contextualized representations that strictly align local protein imputations with holistic cellular phenotypes.

Empowered by this universal and structurally rigorous pre-training, scpFormer serves as a versatile engine for a diverse array of downstream tasks. In the following sections, we demonstrate that the emergent global cell state universally captures high-fidelity, dimension-agnostic representations. These dense embeddings provide a more informative alternative to raw expression profiles, enabling robust cell-type annotation, unsupervised clustering, and massive-scale batch integration across highly discordant multi-center cohorts (Fig. \ref{fig:fig_overview}C). Crucially, we show that the profound protein co-expression logic internalized at the single-cell resolution is highly transferable: by substituting noisy bulk expression inputs with our contextualized latent representations, scpFormer effectively captures complex functional phenotypes to enhance cancer drug response prediction. Finally, we highlight the model's distinct generative capabilities through highly accurate \textit{in silico} panel expansion (cross-panel protein imputation), maximizing the utility of strictly limited experimental panels.

\subsection{scpFormer achieves high accuracy in cell type annotation}

To fine-tune the pretrained scpFormer for cell type annotation, a three-layer neural network classifier takes the scpFormer-derived cell embedding (the [CLS] token embedding) as input to generate categorical predictions for cell types. The whole model was trained with cross-entropy on a reference dataset with expert annotations and then used to predict cell types on a held-out query data partition.

\subsubsection{scpFormer demonstrates strong performance in intra-dataset comparisons}
First, we evaluated the quality of the pre-trained representations on the Levine study~\cite{levine2015data} (Fig.~\ref{fig:fig_cell_type_annotation}A), a commonly used benchmark dataset profiling human bone marrow cells. We applied the fivefold cross-validation strategy to avoid the influence of random results on the conclusion. We visualized the classification performance in Fig. ~\ref{fig:fig_cell_type_annotation}B (Table S2A). Notably, scpFormer achieved exceptional precision for the majority of the 14 immune cell populations over established baselines, including Seurat~\cite{butler2018integrating} and CellTypist~\cite{dominguez2022cross}. As shown in the confusion matrix (Fig.~\ref{fig:fig_cell_type_annotation}C, Table S2B), scpFormer achieved exceptional accuracy ($>$0.95) for the majority of the 14 immune cell populations. The most notable improvements were observed in identifying phenotypically similar subtypes and rare progenitor cells along the continuous hematopoiesis trajectory. Traditional methods had a severe struggle with hematopoietic stem cells (CD34+ CD38lo HSCs) and progenitor populations (CD34 + CD38 + CD123 + HSPCs), obtaining prediction precisions of merely 0.48 and 0.66 in Seurat, respectively (Fig. ~\ref{fig:fig_cell_type_annotation}D, Table S2C). In contrast, scpFormer correctly annotated 0.95 of HSCs and 0.98 of HSPCs. Finally, scpFormer constantly outperformed the Seurat~\cite{butler2018integrating} and CellTypist~\cite{dominguez2022cross} in all classification metrics, including accuracy, precision, recall, and macro F1. Especially, scpFormer achieves a macro F1-score of 0.978 (compared to 0.868 for Seurat~\cite{butler2018integrating} and 0.920 for CellTypist~\cite{dominguez2022cross}). This suggests that scpFormer can handle rare and dominant cell populations with comparable precision, thereby mitigating the bias toward major cell types commonly observed in traditional methods.

\subsubsection{scpFormer consistently performs well in cross-batch cell type annotation}
In real-world clinical and biological applications, references and query datasets are invariably sourced from multiple independent studies, different sequencing platforms, or different donors. Such inherent technical variations often obscure genuine biological signals, leading to catastrophic failures in conventional cell type annotation methods. Therefore, demonstrating cross-batch robustness is a critical prerequisite for any foundation model intended for broad adoption. To evaluate this ability, we benchmarked scpFormer against Seurat~\cite{butler2018integrating} and CellTypist~\cite{dominguez2022cross} using a comprehensive peripheral blood mononuclear cell (PBMC) dataset from Hao et al.~\cite{hao2021integrated} (Fig.~\ref{fig:fig_cell_type_annotation}E). We designed a challenging out-of-distribution evaluation task: the model was fine-tuned exclusively on data from batch 1 (reference) and subsequently tasked with annotating cells from batch 2 (query).

The results demonstrate that scpFormer substantially outperforms established baselines under batch variations (Fig.~\ref{fig:fig_cell_type_annotation}F, Table S3A). As evidenced by the confusion matrices (Fig.~\ref{fig:fig_cell_type_annotation}G, Table S3B), scpFormer maintained high precision even for highly specialized or transitioning cell subsets (e.g., ASDC and intermediate B cells), whose subtle proteomic signatures are often overshadowed by inter-batch variability. While traditional clustering-based integration methods such as Seurat~\cite{butler2018integrating} achieved an accuracy of 0.865, their performance was limited by imbalanced distribution of cell states across batches, yielding a macro F1-score of merely 0.642. CellTypist~\cite{dominguez2022cross} performed even more poorly in this cross-batch scenario (F1-score: 0.577). In contrast, scpFormer achieved an overall accuracy of 0.918 and elevated the macro F1-score to 0.789 (Fig.~\ref{fig:fig_cell_type_annotation}H, Table S3C). These results indicate that, by pre-training on a massive scale, scpFormer learns batch-invariant, intrinsic protein co-expression landscapes rather than memorizing batch-specific technical noise. Therefore, scpFormer functions as a robust pattern-recognition engine, capable of discovering and generalizing cell-type-specific representations, independent of technical confounding factors.

\subsection{Effective cell clustering and batch integration with scpFormer}

A fundamental challenge in single-cell proteomics is the harmonization of datasets between different donors, sequencing platforms, and experimental batches. Conventional data integration algorithms frequently suffer from a severe trade-off: aggressive batch correction often leads to the over-alignment and subsequent loss of fine-grained biological heterogeneity, while conservative correction fails to bridge substantial technical gaps. 

To evaluate whether scpFormer overcomes this bottleneck, we benchmarked the unsupervised latent representations of scpFormer against widely used data integration methods (Harmony~\cite{korsunsky2019fast} and ComBat~\cite{johnson2007adjusting}) on two comprehensive multi-batch datasets: a highly confounded CITE-seq dataset from Zheng et al. study~\cite{zheng2025adtnorm} and a frequently used benchmark dataset from the Stuart et al. study (GSE128639)~\cite{stuart2019comprehensive}. Both datasets contain the cell type and batch information. We quantified the integration quality using the rigorous single-cell integration benchmark (scIB) framework~\cite{luecken2022benchmarking}, which comprehensively evaluates both biological variance conservation (AvgBIO) and batch effect removal (AvgBATCH).

On the Zheng et al. dataset~\cite{zheng2025adtnorm} comprising 6 cell types from 37 batches or studies (Fig. ~\ref{fig:fig_cell_clustering}A\&B, Table S4A), Harmony~\cite{korsunsky2019fast} severely distorted the biological signal in this context, dropping its AvgBIO to 0.457 and ARI to 0.341 (Fig. ~\ref{fig:fig_cell_clustering}C\&D, Table S4B), indicative of catastrophic over-correction. While ComBat~\cite{johnson2007adjusting} provided reasonable clustering, scpFormer achieved best performance, with an AvgBIO of 0.724 and an AvgBATCH of 0.911 (Fig.~\ref{fig:fig_cell_clustering}C\&D, Table S4B). This substantial margin in ASW (Fig.~\ref{fig:fig_cell_clustering}C, Table S4B) demonstrates that while linear methods like ComBat~\cite{johnson2007adjusting} might align dataset means, scpFormer resolves the complex, non-linear batch structures while preserving the crisp boundaries between distinct immunophenotypes.

On the Stuart et al. dataset~\cite{stuart2019comprehensive}(Fig. ~\ref{fig:fig_cell_clustering}E\&F, Table S5A), scpFormer yielded the highest AvgBIO score (0.795 vs. 0.636 for ComBat~\cite{johnson2007adjusting} and 0.625 for Harmony~\cite{korsunsky2019fast})(Fig. ~\ref{fig:fig_cell_clustering}G\&H, Table S5B). Notably, the unsupervised clustering metrics in the scpFormer latent space improved substantially: the Adjusted Rand Index (ARI) increased to 0.833 (compared to Harmony's 0.623), and the Normalized Mutual Information (NMI) also rised to 0.833. Simultaneously, it achieved a high batch integration score (AvgBATCH = 0.957), effectively mixing analogous cell states originating from different experimental batches. These rigorous quantitative metrics suggest that scpFormer does not merely perform local heuristic alignments or linear shifts. Instead, the self-attention mechanism within its transformer architecture effectively decouples technical artifacts from intrinsic biological signatures, projecting highly confounded single-cell proteomic profiles into a unified, biologically faithful embedding space.

\subsection{scpFormer enhances the accuracy of protein expression imputation}

In single-cell proteomics, stochastic dropouts and technical missing data present substantial obstacles that can obscure important molecular signals. Traditional imputation algorithms often rely on local neighborhood smoothing (e.g., k-NN) or linear dependencies, which frequently result in artificial noise or the over-smoothing of rare cell variations. By capturing protein co-expression patterns during extensive pre-training, scpFormer could accurately infer missing protein abundances by leveraging the global contextual dependencies of all observed markers.

\subsubsection{scpFormer achieves accurate protein imputation as validated \textit{in silico}}
To rigorously validate this capability, we first designed an \textit{in silico} imputation benchmark using the Levine dataset~\cite{levine2015data}. We employed a rigorous five-fold cross-validation strategy, systematically masking the expression of eight target proteins to simulate dropout events. We then fine-tuned scpFormer alongside established regression models (random forest, k-NN, and linear regression) to predict the masked values of these specific markers. Evaluating the Pearson correlation coefficient between the predicted and ground-truth protein abundances, scpFormer achieved a overall mean Pearson correlation of 0.751, outperforming random forest (0.711), k-NN (0.685), and linear regression (0.648) (Fig.~\ref{fig:fig_Imputation}A, Table S6A). Fig.~\ref{fig:fig_Imputation}B demonstrated the Pearson Correlation between eight imputed genes with true expression in each fold from each method, scpFormer achieved the best mean of Pearson correlation. Meanwhile, scpFormer demonstrated an exceptional advantage when imputing challenging, low-abundance markers that lack simple linear correlations. For instance, the predictive correlation for FLT3 increased to 0.452 using scpFormer, whereas conventional methods largely failed, yielding correlations near 0.20. This indicates that our attention mechanism reconstructs missing values based on complex, non-linear cellular contexts rather than mere nearest-neighbor averaging.

\subsubsection{Zero-shot imputation of scpFormer restores biological manifolds in clinical cohorts}
Having established its precision against the ground truth, we next investigated scpFormer’s utility in a real-world clinical setting fraught with natural data missingness. We utilized the SARS-CoV-2 multisystem inflammatory syndrome in children (MIS-C) dataset (GSE166489)~\cite{ramaswamy2021immune}. Because the true protein abundances for natural dropouts are inherently inaccessible, we adopted an unsupervised, topology-based evaluation strategy. Notably, we employed scpFormer in a zero-shot manner—directly predicting missing expressions without any dataset-specific fine-tuning—and assessed the biological integrity of the resulting imputed latent space using the scIB framework.

Interestingly, this zero-shot imputation yielded the most biologically coherent cell representations (Fig.~\ref{fig:fig_Imputation}D-F, Table S7). It achieved the highest overall biological conservation score (avgbio = 0.618) and the best clustering agreement (NMI = 0.709, ARI = 0.632) compared to all baseline imputation strategies. Furthermore, scpFormer effectively mitigated latent batch effects, yielding an avgbatch score of 0.912. The corresponding UMAP visualizations confirmed these quantitative gains, revealing that scpFormer effectively rescued the defining signatures of specialized immune subsets that were otherwise blurred by technical dropouts. Together, these results demonstrate that the generalized representations within scpFormer can be universally applied out-of-the-box to restore high-fidelity biological manifolds in sparse clinical cohorts.

\subsection{scpFormer enhances the accuracy of cancer drug response prediction}

To further demonstrate the translational utility of scpFormer, we extended its evaluation to cancer drug response (CDR) prediction (Fig.~\ref{fig:drug_response}A). Accurate estimation of the half-maximal inhibitory concentration ($IC_{50}$) is central to precision oncology, yet remains challenging owing to the pervasive noise and limited concordance between mRNA abundance and drug efficacy. We hypothesized that scpFormer embeddings, by capturing the functional state of the proteome, would provide a more deterministic signal for drug sensitivity than conventional transcriptomic profiles.

To test this hypothesis, we integrated scpFormer into the DeepCDR~\cite{liu2020deepcdr} framework by replacing its standard transcriptomic inputs with scpFormer-derived proteomic embeddings, aligning available bulk protein expression data onto scpFormer's semantic manifold. Predictive performance was subsequently benchmarked against the gene-expression-based scFoundation embedding across multiple drugs and cancer types (Fig.~\ref{fig:drug_response}A).

Compared with scFoundation, scpFormer shows consistently better drug-response prediction performance across these benchmarks. At the global cell line and drug prediction task, scpFormer shows higher Pearson correlation between predicted $IC_{50}$ ($r = 0.921$) than scFoundation ($r = 0.888$) (Fig.~\ref{fig:drug_response}B, Table S8). Lower mean of absolute error (MAE) is observed for more compounds with scpFormer, among 223 tested drugs, 183 drugs have lower MAE from scpFormer than scFoundation. Top ten MAE improved drugs are WH-4-023, A-770041, Refametinib, PD0325901, Epothilone B, AZ628, Tanespimycin, CAY10603, GSK1070916, and Trametinib (Fig.~\ref{fig:drug_response}C), consistent with known functional drug classes such as kinase inhibitors, microtubule-stabilizing agents, and epigenetic modulators. 

Pathway-level stratification further confirmed that scpFormer's gains are not confined to a single therapeutic class; across nearly all drug pathway categories, the majority of compounds exhibited lower MAE with scpFormer compared to scFoundation (Fig.~\ref{fig:drug_response}D). Notably, scpFormer outperformed scFoundation across all compounds in eight pathway categories: p53 pathway, WNT signaling, apoptosis regulation, cell cycle, chromatin histone methylation, cytoskeleton, DNA replication, and ABL signaling, demonstrating consistent advantage in drug response prediction.

\section{Discussion}
In this work, we introduce scpFormer, a generative foundation model that adapts the Transformer architecture to the unique challenges of single-cell proteomics. By shifting from discrete, index-based tokenization from traditional single-cell foundation models to a continuous, sequence-aware representation for addressing the challenges of single-cell proteomics, scpFormer directly targets the fundamental limitation of single-cell proteomics: the fragmentation of protein feature space across technologies, panels, and experimental designs. Our comprehensive analysis demonstrate that leveraging the semantic richness of protein sequences in scpFormer enables the integration of heterogeneous single-cell proteomics experimental platforms and the inference of functional cellular states with fidelity that exceeds transcriptomic proxies, reinforcing the idea that proteomic data more closely reflects cellular phenotypes than transcriptomic profiles and illustrating the potential of scpFormer for interpreting biological phenotypes.. 

A key implication of our work is that directly modeling the proteome at single-cell resolution provides a more functionally grounded and stable representation of cellular states than transcriptomic data. While single-cell RNA sequencing (scRNA-seq) foundation models provide broad genome-scale coverage, mRNA abundance exhibits substantial stochasticity, driven by transcriptional bursting and rapid degradation, which frequently obscures downstream functional consequences. By directly modeling protein co-expression logic, scpFormer buffers this upstream transcriptomic noise. This theoretical advantage is empirically validated by our drug response prediction outcomes. When single-cell derived scpFormer embeddings are utilized to substitute noisy transcriptomic or bulk protein inputs in the DeepCDR framework, we observe an enhancement in predicting cancer drug sensitivity. This cross-scale generalization, from single-cell representations to bulk-level pharmacological responses, provides strong evidence that scpFormer captures a highly stable, functionally grounded signal essential for precision oncology.

Beyond its improved predictive performance relative to many state-of-the-art methods, a key conceptual contribution of scpFormer is its ability to overcome the closed-set constraint that has historically limited both statistical pipelines and Transformer-based approaches in single-cell proteomics analysis. Conventional integration methods are restricted to the intersection of shared features, necessitating the discard of non-overlapping markers. In contrast, scpFormer operates in an open-vocabulary regime, enabling the model to learn from the union of all observed proteins across datasets. The observed monotonic improvement in classification accuracy upon the inclusion of unseen proteins demonstrates that scpFormer captures a generalized grammar of protein co-expression and functional coordination, rather than memorizing fixed protein identities. In this framework, SCP data heterogeneity, traditionally viewed as a liability, becomes a source of effective data augmentation. By anchoring proteins in a continuous evolutionary space via ESM, scpFormer enables zero-shot in silico panel expansion, allowing unmeasured markers to be inferred based on learned biological context. This capability has direct implications for single-cell proteomics experimental design, suggesting that sparse or targeted panels can be computationally enriched without additional wet-lab cost.

Furthermore, the model’s strong performance in massive-scale batch integration highlights the capacity of attention-based architectures to capture global, nonlinear dependencies. Traditional integration tools face a severe trade-off: linear methods (like ComBat) fail to bridge substantial technical gaps, while aggressive alignment methods (like Harmony) often over-correct, obliterating fine-grained biological heterogeneity. scpFormer’s unsupervised disentanglement, as demonstrated across highly confounded CITE-seq and mass cytometry datasets, seamlessly reconciles this conflict. By utilizing its global classification token as an information bottleneck, the model implicitly decouples intrinsic biological signatures from technical nuisance factors—preserving crisp immunophenotypic boundaries while mitigating profound batch effects without requiring explicit, rigid correction objectives.

Despite these advances, several limitations warrant consideration and motivate future work. First, while scpFormer is inherently robust to missing data, it currently treats stochastic dropout implicitly. Incorporating explicit noise models for mass spectrometry dropout could further enhance robustness. Second, the current tokenization strategy relies on canonical amino acid sequences. Consequently, post-translational modifications (PTMs), such as phosphorylation or glycosylation, are not yet explicitly distinguished unless distinct antibodies targeting these states are treated as unique semantic tokens. future extensions incorporating PTM-aware embeddings would enable more accurate modeling of dynamic signaling states. Finally, extending scpFormer to spatially resolved proteomics represents a critical next step, as integrating spatial context with open-vocabulary proteomic representations will be essential for understanding tissue architecture and multicellular organization.

\section{Methods}
\subsection{Pretraining data collection and preprocessing}
\subsubsection{Collection of the single-cell proteomic data}

To construct the single-cell proteomics pre-training database, we performed a systematic search of PubMed, GEO, and the PRIDE Archive using single-cell proteomics-related terminology, followed by manual curation to verify data availability and quality. This collection was further supplemented with datasets sourced from two dedicated single-cell proteomics repositories, SingPro~\cite{lian2024singpro} and SPDB~\cite{wang2024spdb}. All retrieved datasets were uniformly preprocessed and filtered (Table S1). Following standardized preprocessing and integration, the final corpus comprises 391,280,332 single-cell measurements restricted to human samples.
The curated dataset collectively encompasses 544 proteins quantified spanning 22 distinct cell types (Table S1). The two platform categories exhibit complementary characteristics: antibody-based technologies afford broad cellular coverage at large scale, whereas mass spectrometry-based approaches provide substantially greater proteomic depth per cell. Representative studies and corresponding cell counts stratified by technology are summarized in Table S1. Collectively, this database constitutes a large-scale, heterogeneous corpus suitable for developing a foundation model capable of generalizing across diverse antibody panels, sparse measurement profiles, and platform-specific technical variation.
\subsubsection{Protein symbol unification}
Protein symbols across all raw count expression matrices were standardized by mapping each entry to its canonical identifier using the UniProt reference database. The corresponding amino acid sequences were subsequently retrieved from UniProt for downstream structural and functional encoding.

\subsection{Tokenizer and cell representations}
Formally, let $\mathbf{X} \in \mathbb{R}^{N \times M}$ denote the single-cell proteome expression matrix, where $N$ is the total number of cells and $M$ represents the global universe of distinct protein features across all panels. A cell $i$ is naturally represented as a vector $\mathbf{x}^{(i)} \in \mathbb{R}^{M}$, containing measured expression values for a set of proteins. Unlike natural language processing tasks where the input is a sequence of discrete words, single-cell proteomic data exhibits a dual modality: each measured feature inherently carries both a semantic identity (i.e., ``Which protein is this?'') and a continuous scalar magnitude (i.e., ``What is its expression level?''). To optimally capture this distinct dual modality, we introduce a single-cell proteomics-specific tokenization strategy to encode individual cells. These tokenized sequences are subsequently processed by a self-attention based Transformer encoder to extract a universal, dimension-agnostic representation for each cell, which serves as the foundation for all downstream analytical tasks. The details are described in the following section.

\subsubsection{Protein tokenizer}
Standard Transformer architectures typically rely on a fixed-vocabulary paradigm, utilizing a randomly initialized lookup table for a static set of discrete tokens. However, this architecture is fundamentally ill-suited for single-cell proteomics, where antibody panels are highly variable across datasets, and a universal, closed-set reference panel does not exist. To overcome this limitation, we introduce a continuous semantic tokenization mechanism. Instead of treating proteins as arbitrary categorical indices, we ground our model in the physicochemical and evolutionary reality of protein sequences. For every protein $p$ present in the dataset, we map its identifier to the corresponding UniProt ID and retrieve its canonical amino acid sequence $S_p = (a_1, a_2, \dots, a_L)$, where $a_k$ represents the $k$-th amino acid residue and $L$ represents the number of amino acid. To capture the structural and functional semantics of these sequences, we leverage the pre-trained Evolutionary Scale Modeling (ESM) framework. For protein $j$ in cell $i$, we compute a dense identity vector $\mathbf{e}_{p,j}^{(i)} \in \mathbb{R}^{d}$ by averaging the token-level output embeddings from the ESM model:
\begin{equation}
\mathbf{e}_{p,j}^{(i)} = \frac{1}{L} \sum_{k=1}^{L} \text{ESM}(a_k)
\end{equation}
By leveraging the pre-trained evolutionary knowledge of ESM, functionally similar proteins are mapped to proximal points in the latent manifold. This expanding vocabulary approach allows the model to generalize zero-shot to new proteins unseen during training.

\subsubsection{Expression value tokenizer}
Single-cell proteomics captures critical physiological gradients where subtle shifts in protein abundance (expression) dictate cellular state transitions. Applying artificial quantization or discrete binning on proteomic data severely compresses the dynamic range and obliterates critical, continuous phenotypic signals. Therefore, we directly encode the raw, continuous expression values. For a non-zero expression value $x_{j}^{(i)} \in \mathbb{R}$ of protein $j$ in cell $i$, we project this scalar magnitude into a high-dimensional dense vector space that matches the dimensionality $d$ of the identity embedding. This is achieved using a multi-layer perceptron (MLP) to capture potential non-linear dynamics in protein abundance:
\begin{equation}
\mathbf{e}_{v,j}^{(i)} = \text{MLP}(x_{j}^{(i)})
\end{equation}
This continuous value embedding explicitly preserves the subtle quantitative differences in protein expression across individual cells without relying on artificial discretization boundaries.

\subsubsection{Cell representation}
The final input representation $\mathbf{h}_j^{(i)}$ for a specific protein $j$ in cell $i$ is constructed by fusing its semantic identity with its measured abundance:
\begin{equation}
\mathbf{h}_j^{(i)} = \mathbf{e}_{p,j}^{(i)} + \mathbf{e}_{v,j}^{(i)}
\end{equation}
To form the complete input sequence $\mathbf{H}^{(i)}$ for the Transformer encoder, we prepend a special learnable classification token ($\mathbf{h}_{cls}$) to aggregate global cellular state information:
\begin{equation}
\mathbf{H}^{(i)} = [\mathbf{h}_{cls}, \mathbf{h}_1^{(i)}, \mathbf{h}_2^{(i)}, \dots, \mathbf{h}_{|\mathcal{P}_i|}^{(i)}]
\end{equation}
Crucially, we construct this sequence by strictly encoding only the proteins that are experimentally quantified in that specific cell. Here, $\mathcal{P}_i$ denotes this exact set of detected proteins for cell $i$, and $|\mathcal{P}_i|$ represents the resulting dynamic sequence length. By doing so, each individual cell is naturally transformed into a variable-length tokenized sequence of its own expressed proteome. This dynamic set construction strategy directly circumvents the missing problem inherent to integrative proteomics and enables our model to process variable-length inputs. Consequently, this architecture enables the seamless integration of multi-center datasets with highly disjoint antibody panels or varying mass spectrometry detection depths, effectively maximizing the utility of all available proteomic information.

The constructed continuous sequence $\mathbf{H}^{(i)}$ is then processed by a multi-layer self-attention Transformer encoder. The output of the Transformer is a sequence of contextualized latent representations:
\begin{equation}
\mathbf{Z}^{(i)} = {Transformer}(\mathbf{H}^{(i)}) = [\mathbf{z}_{cls}, \mathbf{z}_1^{(i)}, \mathbf{z}_2^{(i)}, \dots, \mathbf{z}_{|\mathcal{P}_i|}^{(i)}]
\end{equation}
Here, each output token $\mathbf{z}_j^{(i)}$ represents the final embedding of protein $j$ in cell $i$, updated via self-attention to encode its local co-expression interactions. Complementing these local protein-level representations, the updated classification token $\mathbf{z}_{cls}$ serves as the universal, dimension-agnostic representation of the holistic cellular state. Once pre-training is complete, this universal embedding $\mathbf{z}_{cls}$ acts as the primary information bottleneck. It can be directly extracted for diverse downstream analytical tasks without requiring task-specific architecture modifications.

\subsection{Pre-training}
The pre-training objective of our scpFormer model is fundamentally driven by a critical real-world demand in single-cell proteomics: the in silico imputation and panel expansion of unmeasured proteins based on a strictly limited subset of measured features. To achieve this goal, we introduce a masking strategy specifically tailored to the unique characteristics of single-cell proteomics data, combined with a set of dedicated loss functions. We detail these components below.

\subsubsection{Context-Target masking}
During pre-training, we simulate the real-world scenario of targeted proteomics by partitioning the input proteins of a given cell into two disjoint sets: a context set $\mathcal{P}_{ctx}$ (representing the visible, measured reference panel) and a target set $\mathcal{P}_{tgt}$ (representing the unmeasured proteins to be imputed). To ensure the model learns true conditional inference rather than artifactual memorization, the self-attention mechanism must be strictly constrained. In a real-world panel expansion, unmeasured proteins cannot provide information to one another. Therefore, we introduce an attention mask $\mathbf{M} \in \{0, -\infty\}^{|\mathcal{P}_i| \times |\mathcal{P}_i|}$ designed such that target proteins can only attend to the observed context $\mathcal{P}_{ctx}$ and themselves, but are strictly prohibited from attending to other target proteins. Formally, the attention weight modifier $m_{u,v}$ between query position $u$ and key position $v$ is:
\begin{equation}
m_{u,v} = 
\begin{cases} 
-\infty, & \text{if } u \neq v \text{ and } v \in \mathcal{P}_{tgt}  \\
0, & others \\ 
\end{cases}
\end{equation}
By adding $\mathbf{M}$ in self-attention, the Transformer processes the proteins as an unordered set lacking sequential causality, performing parallel protein imputation conditioned exclusively on the observed context. Based on this masking strategy, we optimize three synergistic loss functions.

\subsubsection{Pre-training Objectives}
To explicitly align our mathematical objectives with the architectural components outlined in our framework overview (Fig. \ref{fig:fig_overview}B), we formulate a hierarchical pre-training framework consisting of three specialized decoding branches. Each decoder tackles the imputation of unmeasured proteins from a distinct biological scale, jointly optimizing the foundation model.

Expression Self-Decoder ($\mathcal{L}_{self}$).
Our first decoding branch focuses on micro-level protein-protein interactions, relying on the contextualized protein embeddings to predict the target expressions. For the masked target subset $\mathcal{P}_{tgt}$, the model infers their continuous expression values based strictly on the unmasked contextual proteins $\mathcal{P}_{ctx}$. We apply a multilayer perceptron (MLP) to the output token embedding $\mathbf{z}_j^{(i)}$ to regress the target expression. Specifically, we compute the Mean Squared Error (MSE) loss between the model prediction and the ground truth continuous magnitude $x_{i,j}$:
\begin{equation}
    \mathcal{L}_{self} = \frac{1}{|\mathcal{P}_{tgt}|} \sum_{j \in \mathcal{P}_{tgt}} \left( MLP(\mathbf{z}_j^{(i)}) - x_{i,j} \right)^2
\end{equation}

Global Expression Decoder ($\mathcal{L}_{global}$). Beyond protein interactions, deriving a highly informative, universal cell representation ($\mathbf{z}_{\text{CLS}}$) is crucial for downstream analytical tasks such as cell typing and batch integration. To ensure this global token does not merely act as a placeholder but actively compresses the entire cellular state, we introduce a consistency objective. This loss forces the model to predict the missing target proteins using only the global cell embedding queried by the target's semantic identity.
Specifically, for each target protein $j$ in cell $i$, we project its protein embedding $\mathbf{e}_{p,j}^{(i)}$ into a dedicated query vector $\mathbf{q}_j^{(i)}$ via a multi-layer perceptron. The predicted continuous expression value $\hat{x}_{i,j}$ is then computed through a parameterized inner product between this query vector and the global cell representation $\mathbf{z}_{cls}$:
\begin{equation}
\begin{aligned}
\mathbf{q}_j^{(i)} = MLP(\mathbf{e}_{p,j}^{(i)}) \\
\hat{x}_{i,j} = {\mathbf{q}_j^{(i)}}^\top \mathbf{W} \mathbf{z}_{cls}
\end{aligned}
\end{equation}
where $\mathbf{W}$ is a learnable weight matrix mapping the cell state space to the protein query space. We then compute the Mean Squared Error (MSE) loss between this inner-product prediction and the ground truth continuous magnitude $x_{i,j}$:
\begin{equation}
\mathcal{L}_{global} = \frac{1}{|\mathcal{P}_{tgt}|} \sum_{j \in \mathcal{P}_{tgt}} \left( \hat{x}_{i,j} - x_{i,j} \right)^2
\end{equation}
This parameterized querying mechanism serves as a low-rank information bottleneck, forcing $\mathbf{z}_{\text{CLS}}$ to anchor a robust, batch-invariant biological state that intrinsically aligns with the physicochemical identities of the proteins it aims to predict.

Joint Decoder ($\mathcal{L}_{joint}$). While $\mathcal{L}_{self}$ effectively imputes missing values based on protein-protein interactions, it operates from a ``cold start'' where the initial classification token lacks specific cellular context. To explicitly bridge the micro-level protein network with the macro-level cellular phenotype, we implement a state-conditioned refinement phase. In this generative forward pass, the model processes the exact same input sequence (context and target proteins), but we explicitly replace the generic, uninformative [CLS] token with the fully contextualized, detached global cell embedding ($\mathbf{z}_{cls}^{detach}$) learned from the initial forward pass. Formally, we construct a refined input sequence $\widetilde{\mathbf{H}}^{(i)}$ for cell $i$:
\begin{equation}
\widetilde{\mathbf{H}}^{(i)} = [\mathbf{z}_{\text{CLS}}^{detach}, \mathbf{h}_1^{(i)}, \mathbf{h}_2^{(i)}, \dots, \mathbf{h}_{|\mathcal{P}_i|}^{(i)}]
\end{equation}
This sequence is fed back into the shared Transformer encoder to produce the refined, state-conditioned latent embeddings:
\begin{equation}
\widetilde{\mathbf{Z}}^{(i)} = \text{Transformer}(\widetilde{\mathbf{H}}^{(i)})
\end{equation}
The refined continuous expression prediction for each target protein $j$ is then decoded via a shared MLP layer. Finally, we compute the Mean Squared Error (MSE) against the ground truth $x_{i,j}$:
\begin{equation}
    \mathcal{L}_{joint} = \frac{1}{|\mathcal{P}_{tgt}|} \sum_{j \in \mathcal{P}_{tgt}} \left( MLP(\widetilde{\mathbf{z}}_j^{(i)}) - x_{i,j} \right)^2
\end{equation}

By injecting the mature global cellular state back into the sequence as a prior, $\mathcal{L}_{SGR}$ forces the self-attention mechanism to re-evaluate the local protein imputations. This acts as a powerful structural regularizer, ensuring that the inferred expression of unmeasured target proteins is strictly concordant with the holistic physiological state of the cell, thereby preventing local imputation artifacts common in highly sparse data. 

All three decoder objectives are computed and jointly optimized within a single training iteration. The final loss is:
\begin{equation}
    \mathcal{L}_{final} = \mathcal{L}_{self}+\mathcal{L}_{global}+\mathcal{L}_{joint}
\end{equation}

\subsection{Downstream tasks}
\subsubsection{Annotation}
For the cell type-annotation task, we fine-tuned the model on a reference set with ground truth labels and validated annotation performance on a held-out query set. The common set of proteins between the pretrained foundation model and the reference set was retained. All pretrained model weights were used to initialize the fine-tuned model, except for the output cell type classifier, which was randomly initialized. The cell type-classification fine-tuning objective was used to minimize the classification loss. Specifically, the universal cell representation $\mathbf{z}_{cls}^{(i)}$ is fed into a multi-layer perceptron classifier to predict the probability distribution $\hat{\mathbf{y}}^{(i)}$ over $K$ predefined cell types. The entire network is optimized end-to-end using the standard cross-entropy loss $\mathcal{L}_{CE}$ against the ground truth one-hot encoded labels
$\mathbf{y}^{(i)}$:
\begin{equation}
\begin{aligned}
    \hat{\mathbf{y}}^{(i)} = \text{Softmax}(MLP(\mathbf{z}_{cls}^{(i)}))\\
    \mathcal{L}_{CE} = - \sum_{c=1}^{K} y_c^{(i)} \log(\hat{y}_c^{(i)})
\end{aligned}
\end{equation}

\subsubsection{Clustering and integration}
Batch effects confound the integration of single-cell proteomic datasets across different experimental runs or platforms. To extract the most comprehensive global cell representation ($\mathbf{z}_{cls}$) for this task, we replace the restricted context-target masking used in pre-training with bidirectional attention, allowing the classification token to leverage the full context of shared proteins across batches. Alongside the pre-training reconstruction ($\mathcal{L}_{CPR}$) and consistency ($\mathcal{L}_{GPC}$) objectives, we jointly optimize two integration-specific losses to explicitly correct batch artifacts and preserve biological variance: 

Domain Adaptation via Gradient Reversal ($\mathcal{L}_{DAR}$). To regress out non-biological batch variations, we employ adversarial domain adaptation. A Gradient Reversal Layer (GRL) connects the global output $\mathbf{z}_{CLS}^{(i)}$ to a MLP-based batch-prediction classifier. We optimize the adversarial cross-entropy loss:
\begin{equation}
\mathcal{L}_{DAR} = \mathcal{L}_{CE}(MLP(\text{GRL}(\mathbf{z}_{cls}^{(i)})), b^{(i)})
\end{equation}
where $b^{(i)}$ is the batch label. The GRL reverses gradients during backpropagation, forcing the Transformer encoder to project cells into a batch-invariant latent manifold.

Elastic Cell Similarity ($\mathcal{L}_{ECS}$). To enhance the topological separability of the integrated manifold, we apply a contrastive Elastic Cell Similarity loss:$$\mathcal{L}_{ECS} = \frac{1}{|\mathcal{B}|^2} \sum_{i, i' \in \mathcal{B}} -\left( \cos(\mathbf{z}_{cls}^{(i)}, \mathbf{z}_{cls}^{(i')}) - \beta \right)^2$$where $\cos(\cdot, \cdot)$ denotes the cosine similarity between cells within a mini-batch $\mathcal{B}$, and $\beta$ is a predefined elasticity threshold. This objective explicitly amplifies the similarity of related cells while pushing distant cells further apart, refining cluster boundaries.

\subsubsection{Imputation}
For the missing protein imputation task, the objective is to accurately reconstruct the continuous expression values of unmeasured proteins (due to technical dropouts or disjoint panels) based on a limited observed reference panel. To achieve this, we directly adapt the $\mathcal{L}_{self}$ objective utilized during pre-training.

\subsubsection{Drug responses}
To evaluate the predictive power of our universal cell representation in pharmacological contexts, we applied our foundation model to predict cancer cell line drug responses. We utilized paired cell line proteomic profiles and corresponding drug sensitivity data, specifically the half-maximal inhibitory concentration ($\text{IC}_{50}$). Rather than relying on raw, highly sparse expression matrices, we processed the continuous protein expression profile of each cell line through our pre-trained encoder to extract the robust global state embedding ($\mathbf{z}_{cls}$). We then integrated these dense, context-rich embeddings into the established DeepCDR framework~\cite{liu2020deepcdr}, directly substituting the baseline raw expression inputs. The downstream model was trained to predict the $\text{IC}_{50}$ values, and the predictive performance was quantitatively evaluated using the Pearson Correlation Coefficient (PCC) between the model predictions and the ground truth $\text{IC}_{50}$ across varying cell lines and drug compounds.

\subsection{Implementation details}

To stabilize training across the highly variable dynamic ranges of single-cell protein abundances, raw continuous expression matrices were first transformed using either a $\log(1+x)$ or $\text{arcsinh}$ function, followed by Min-Max normalization strictly scaled to a continuous range of $[0, 10]$. The core foundation model is instantiated with 12 stacked Transformer encoder blocks, utilizing a hidden embedding dimension of $d=512$ and 8 attention heads. To maximize computational efficiency and accommodate highly multiplexed protein sequences, we integrated FlashAttention-2 into the self-attention mechanism, significantly reducing memory complexity.  During the masked modeling phase, the masking ratio (the proportion of target proteins to generate) was dynamically and uniformly sampled from $\{0.15, 0.30, 0.45\}$ for each sequence. This dynamic masking strategy exposes the model to varying degrees of data sparsity, enhancing its robustness to severe dropouts. Distributed pre-training was executed in a Python 3.10 environment on a single compute node equipped with 8 NVIDIA H100 GPUs. By allocating a massive per-GPU batch size of 768 cells, we achieved an effective global batch size of 6,144 cells per step, inherently stabilizing the gradient estimation. The network was optimized using the AdamW optimizer with an initial learning rate of $1 \times 10^{-4}$. To prevent early representational collapse, we employed a linear learning rate warmup for the first 10,000 steps, followed by a cosine annealing decay schedule. To ensure strict experimental reproducibility, all computational environments and stochastic initializations were secured with a global random seed of 42.

\section{Data availability}
Levine et al.~\cite{levine2015data} dataset was accessed from \url{https://github.com/lmweber/benchmark-data-Levine-32-dim}. The Hao et al.~\cite{hao2021integrated} dataset is publicly accessible from the GEO database using accession number \href{https://www.ncbi.nlm.nih.gov/geo/query/acc.cgi?acc=GSE164378}{GSE164378}. Zheng et al. dataset~\cite{zheng2025adtnorm} was retrieved from \url{https://github.com/yezhengSTAT/ADTnorm}. Stuart et al.~\cite{stuart2019comprehensive} dataset is accessible from the GEO database via accession number \href{https://www.ncbi.nlm.nih.gov/geo/query/acc.cgi?acc=GSE128639}{GSE128639}. Ramaswamy et al. dataset (MIS-C) is public in \href{https://www.ncbi.nlm.nih.gov/geo/query/acc.cgi?acc=GSE166489}{GSE166489}. 

\bibliography{sn-bibliography}
\newpage
\section{Figures}

\begin{figure}[htbp]
  \centering
  \includegraphics[width=\textwidth]{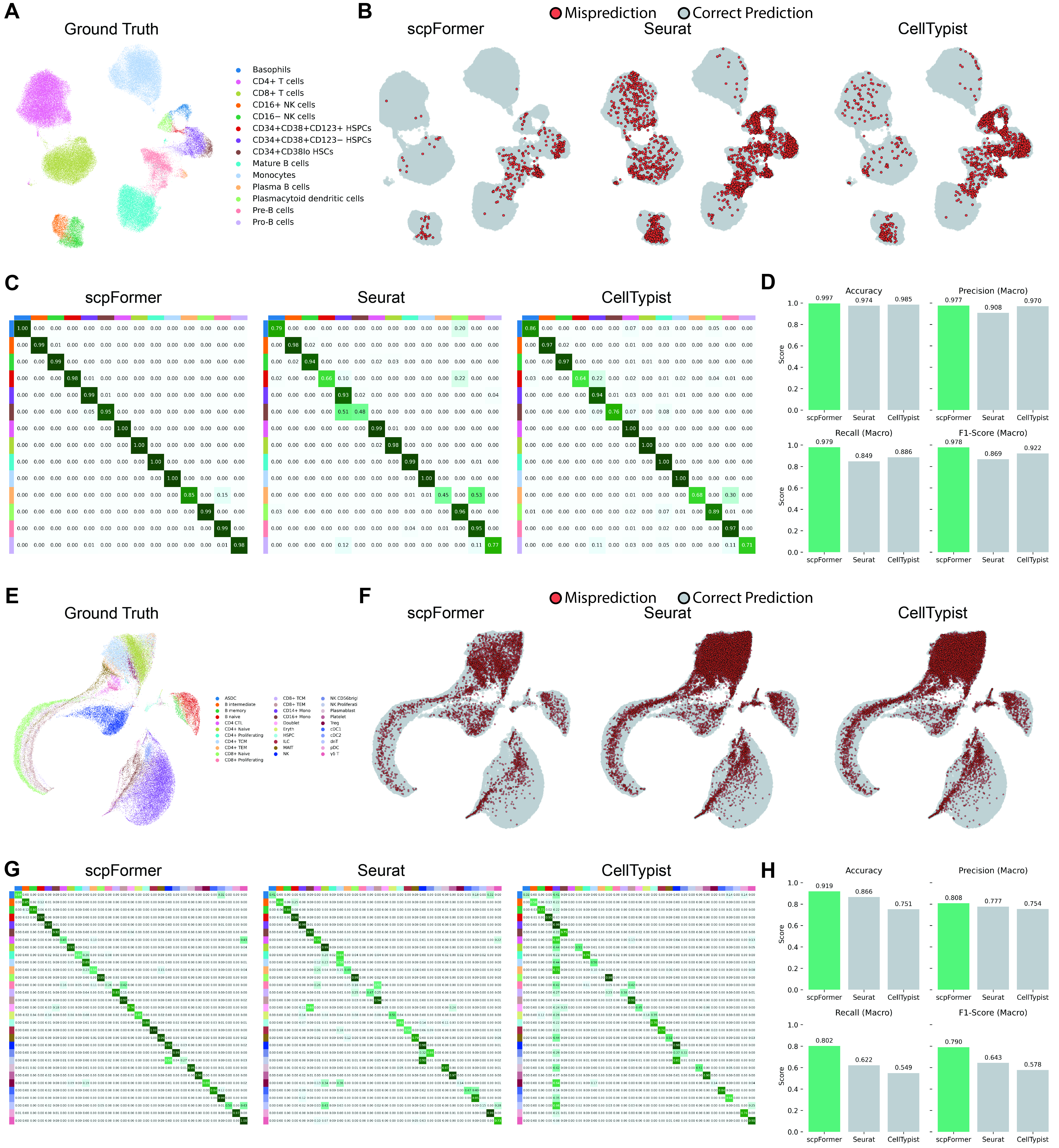} 
  \caption{\textbf{Cell type annotation task.} \textbf{A,E,} UMAP embeddings colored by ground truth cell type annotations for two independent, highly heterogeneous single-cell proteomics datasets: Levine et al. study~\cite{levine2015data} \textbf{A} and Hao et al. study~\cite{hao2021integrated} \textbf{E}. \textbf{B,F,} Spatial distribution of classification errors visualized on the UMAP manifolds. Individual cells are colored based on annotation outcomes (Grey: correct prediction; Red: misprediction) for scpFormer compared against established baseline methods (Seurat and CellTypist). scpFormer exhibits a significant reduction in misprediction density, effectively resolving ambiguous decision boundaries that confound standard algorithms. \textit{(Continued on the following page)}}
\end{figure}

\clearpage
\begin{figure}[!t]
  \ContinuedFloat
  \caption{(Continued) \textbf{C,G,} Confusion matrices detailing the cell-type-specific prediction accuracies for scpFormer, Seurat, and CellTypist. The diagonal values represent the true positive rate for each corresponding subpopulation. scpFormer robustly maintains a strong diagonal, demonstrating strong discriminative performance, particularly for rare or phenotypically subtle cell subsets where baseline methods exhibit severe cross-classification. \textbf{D,H,} Quantitative benchmarking of overall classification performance. scpFormer consistently achieves the highest scores across all rigorous evaluation metrics, including Accuracy, Macro-Precision, Macro-Recall, and Macro F1-Score, in both Levine study~\cite{levine2015data} \textbf{D} and Hao et al. study~\cite{hao2021integrated} \textbf{H}, validating the effectiveness of its contextualized latent representations for precise phenotype inference.}
  \label{fig:fig_cell_type_annotation}
\end{figure}

\begin{figure}[htbp]
  \centering
  \includegraphics[width=\textwidth]{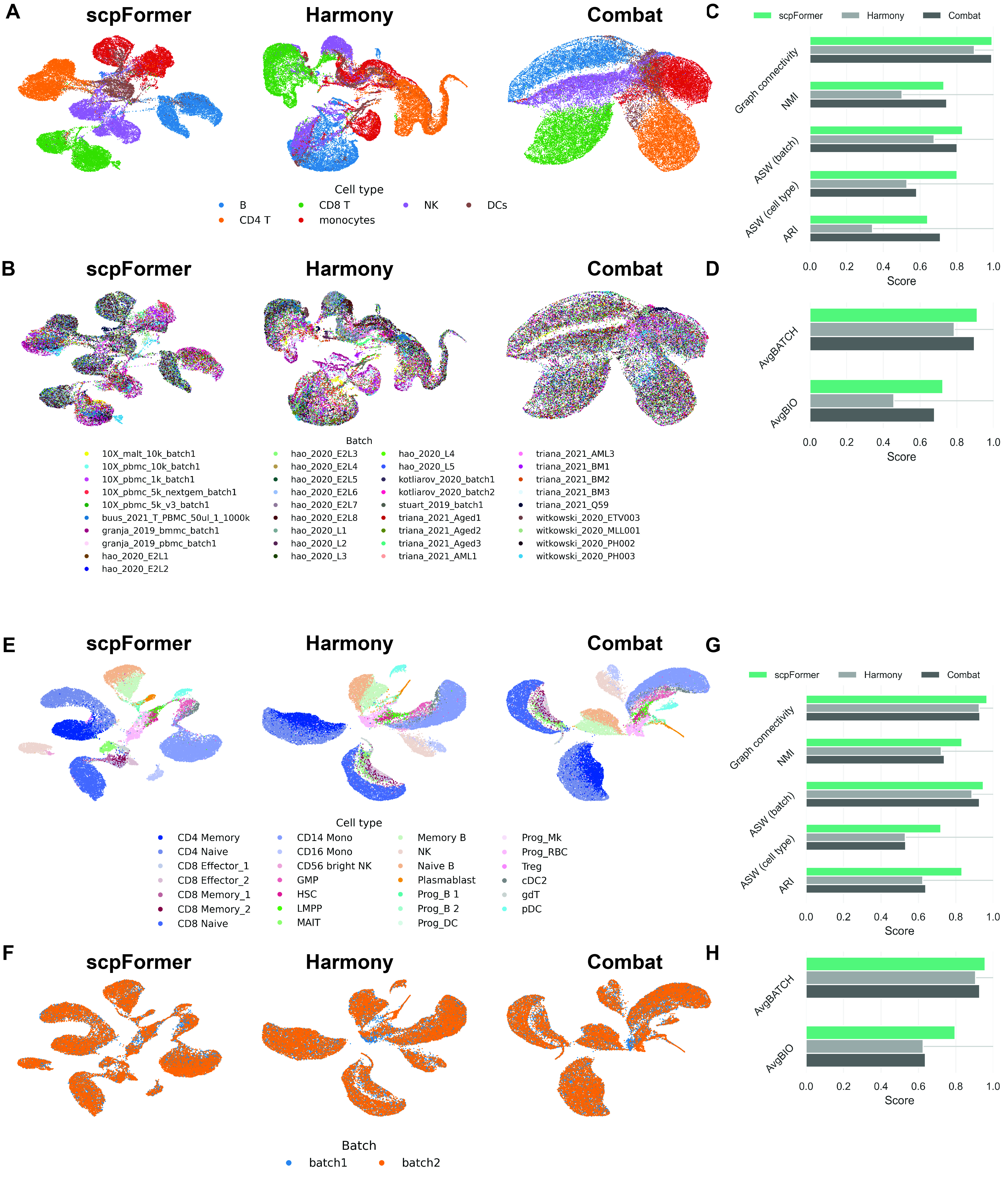} 
  \caption{\textbf{Cell clustering and batch integration tasks.} \textbf{A,B,} UMAP of Zheng et al. study~\cite{zheng2025adtnorm} integrated by scpFormer, Harmony, and ComBat. Cells are colored by ground-truth cell type annotations \textbf{A} and experimental batch origins \textbf{B}. While Harmony exhibits severe over-correction (blurring distinct biological boundaries) and ComBat fails to completely merge batch-specific clusters, scpFormer effectively unifies disparate batches into a shared manifold while maintaining crisp, highly resolved immunophenotypic clusters. \textit{(Continued on the following page)}  }
\end{figure}

\begin{figure}[!t]
  \ContinuedFloat 
  \caption{(Continued) \textbf{C,} Detailed quantitative evaluation of integration performance for Zheng et al. study~\cite{zheng2025adtnorm} using the single-cell integration benchmark (scIB) framework. Metrics include Adjusted Rand Index (ARI), Normalized Mutual Information (NMI), and cell-type Average Silhouette Width (ASW) for biological conservation, alongside batch ASW and Graph connectivity for batch correction. scpFormer substantially outperforms all baselines across these rigorous metrics. \textbf{D,} Aggregated overall scores for Batch Correction and Biological Preservation in Zheng et al. study~\cite{zheng2025adtnorm}. scpFormer effectively breaks the conventional integration trade-off, achieving the highest scores in both dimensions simultaneously.\textbf{E,F,} UMAP for Stuart et al. study~\cite{stuart2019comprehensive}, colored by cell type \textbf{E} and batch \textbf{F}. scpFormer demonstrates effective unsupervised disentanglement, accurately aligning analogous rare cell states (e.g., specific memory or progenitor subsets) across batches without artificial distortion. \textbf{G,H,} Corresponding scIB individual metrics \textbf{G} and aggregated summary scores \textbf{H} for Stuart et al. study~\cite{stuart2019comprehensive}. The sustained performance advantage of scpFormer highlights the robustness and generalizability of its attention-based representations across distinct experimental designs and technical modalities.}
  \label{fig:fig_cell_clustering}
\end{figure}

\begin{figure}[htbp]
  \centering
  \includegraphics[width=\textwidth]{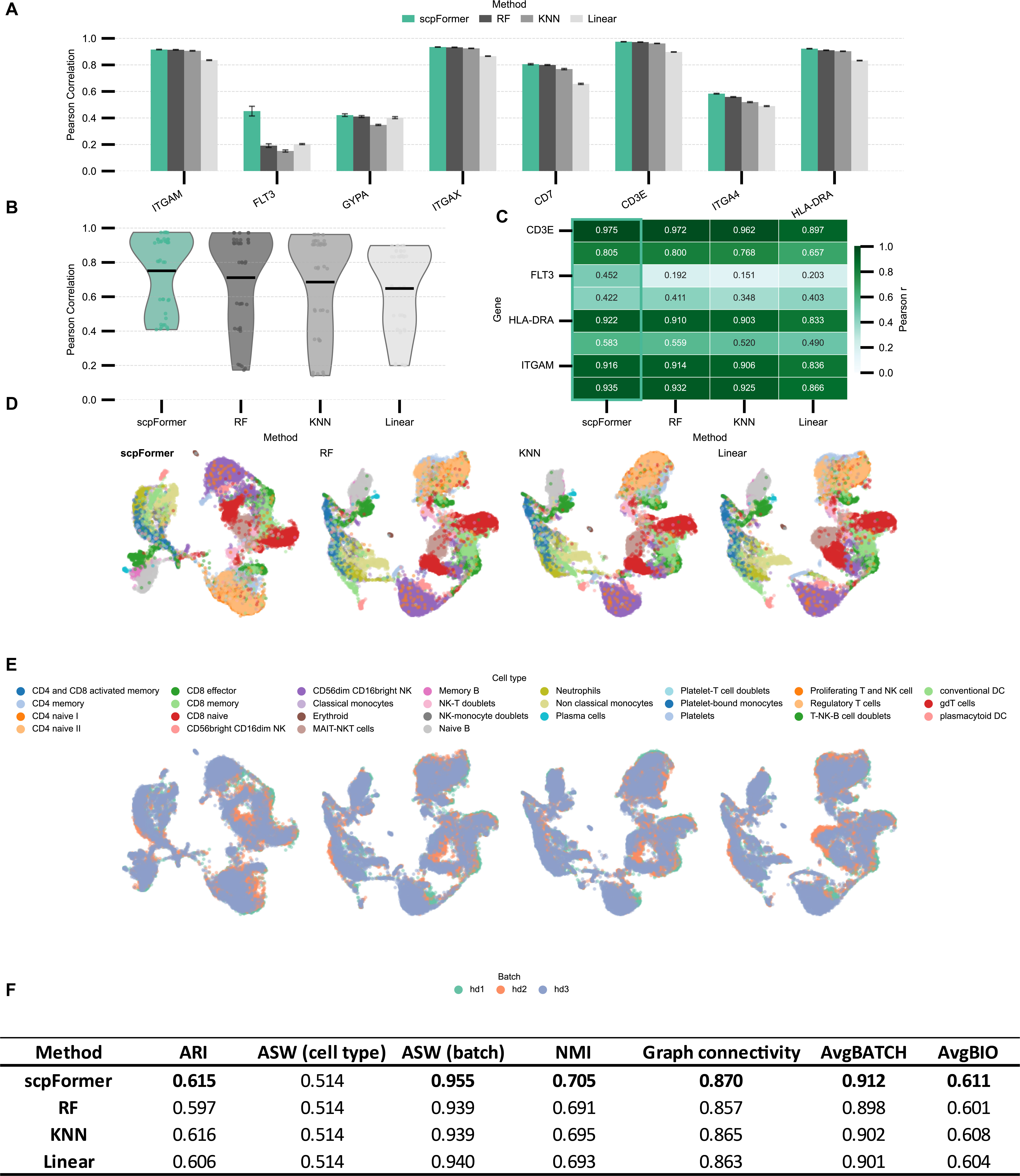} 
  \caption{\textbf{Imputation task}. \textbf{A,} Quantitative evaluation of targeted protein imputation via 5-fold cross-validation in the Levine dataset~\cite{levine2015data}. Bar plots display the Pearson correlation coefficients between predicted and ground-truth expression levels for representative markers. scpFormer significantly outperforms baseline methods (random forest, k-NN, and linear regression), particularly on non-linear or low-abundance features. Error bars represent standard deviation across cross-validation folds. \textbf{B,} Violin plots illustrating the global distribution of Pearson correlation scores across all masked proteins. scpFormer yields a substantially higher overall predictive accuracy and a denser top-tier distribution compared to conventional algorithms. \textit{(Continued on the following page)}}
\end{figure}

\begin{figure}[!t]
  \ContinuedFloat
  \caption{(Continued) \textbf{C,} Detailed heatmap highlighting the accurate prediction correlations for key immunological markers. Notably, scpFormer achieves a high correlation for the FLT3 marker (0.452), whereas local-neighborhood and linear baselines largely fail to reconstruct its complex cellular context. \textbf{D,E,} Unsupervised evaluation of zero-shot imputation in the MIS-C clinical cohort~\cite{ramaswamy2021immune} fraught with natural data missingness. UMAP embeddings generated from the imputed latent spaces are colored by ground-truth cell type annotations \textbf{D} and experimental batch \textbf{E}. Without any dataset-specific fine-tuning, scpFormer uniquely restores crisp immunophenotypic boundaries for specialized subsets while seamlessly mitigating severe latent batch effects. \textbf{F} Comprehensive topology-based benchmarking of the imputed MIS-C manifolds~\cite{ramaswamy2021immune} using the scIB framework. scpFormer universally achieves the highest quantitative scores in both biological variance conservation (NMI, ARI, ASW, and Avg Bio) and batch effect removal (Graph Conn and Avg Batch), validating its capacity to rescue fragmented features out-of-the-box.}
  \label{fig:fig_Imputation}
\end{figure}

\begin{figure}[htbp]
  \centering
  \includegraphics[width=\textwidth]{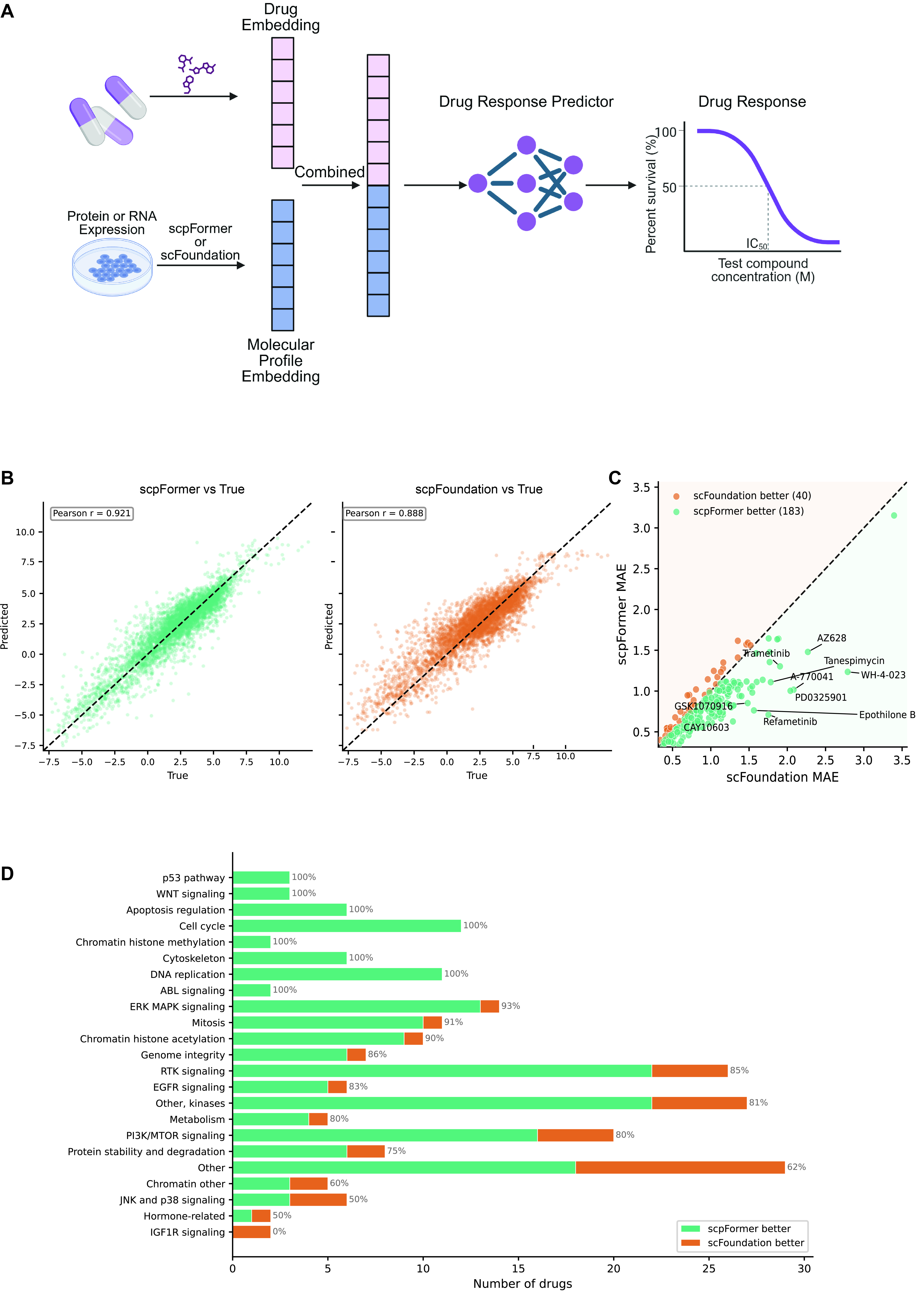} 
  \caption{\textbf{Drug response prediction tasks} \textbf{A,} Schematic of the $IC_{50}$ prediction workflow. For each cell line, drug embeddings and protein expression embeddings are concatenated and passed as input to a neural network for $IC_{50}$ prediction.\textit{(Continued on the following page)}}
\end{figure}

\begin{figure}[!t]
  \ContinuedFloat
  \caption{(Continued) \textbf{B,} Scatter plot of predict $IC_{50}$ and true $IC_{50}$ based on scpFormer and scFoundation method. $r$ represent the increased Pearson's correlation from baseline. \textbf{C,} Drug MAE comparison between scpFormer and scFoundation. Each dot represents a drug and the green colored dots represent the drug gets better prediction from scpFormer than scFoundation. \textbf{D,} Win count summary by drug pathway. Each horizontal bar is one target pathway category. The bar is stacked into two parts, scpFormer better and scFoundation better. The total bar length is total number of evaluated drugs in that pathway. The percentage label at the end is the scpFormer win rate in that pathway.}
  \label{fig:drug_response}
\end{figure}
\end{document}